\documentstyle[prb,aps,preprint,epsf,amssymb,epsfig]{revtex}
\begin{document}
\relax
\begin{titlepage}
\title{ 
  Effective Interactions Between Rigid Polyelectrolytes and \\
  Like-charged Planar
  Surfaces
}
\author{R. Jay Mashl\footnote{Present address: NCSA, University of Illinois,
405 N. Mathews, Urbana, IL  61801; mashl@ncsa.uiuc.edu} and Niels Gr{\o}nbech-Jensen\footnote{ngj@lanl.gov}}
\address{
Theoretical Division and the Center for Nonlinear Studies, MS
  B258, \\
  Los Alamos National Laboratory, Los Alamos, NM~~87545
}
\vskip6pt
\date{Final version published in {\it The Journal of Chemical Physics},
{\bf 109}, 4617--4623 (1998)}

\maketitle

\begin{abstract} 

\setlength{\baselineskip}{0.5\baselineskip}

We study the effective interaction between a planar array of uniformly
negatively charged, stiff rods parallel to a negatively charged planar
substrate in the absence of salt in a continuous, isotropic dielectric
medium.  Using Brownian dynamics simulations, we examine the general
effects of counterion valence, rod spacing, macroion charge densities,
and the rod size on the attractive rod-surface interaction force.  At
room temperature divalent as well as monovalent counterions mediate an
interaction that can be repulsive or attractive upon adjusting either
the macroion charge densities or the rod radius.  Finally, we examine
the effects of discretizing the surface charge as laterally mobile
monovalent anions and of electrostatic images in the substrate.

\vskip12pt

\end{abstract}

\thispagestyle{empty}
\end{titlepage}

\setlength{\baselineskip}{0.5\baselineskip}

\section{Introduction}

In the past few decades many investigations have provided evidence
that effective attractions can occur between like-charged macroions.
Attraction between like-charged surfaces or rod-like macroions has
been observed numerically
\cite{Guldbrand,sven,bratko,guldnils,akesson,valleau,kjell_ake,bolhuis,lyub,us}
mainly through the use of Monte Carlo methods,
\cite{Guldbrand,sven,bratko,guldnils,akesson,valleau,kjell_ake,bolhuis,lyub}
but also through Brownian dynamics simulations.\cite{us} Most of these
studies have featured ions as mediating the attraction, but a few
studies \cite{akesson,podgornik2} have found an attraction by the use
of polyelectrolytes of the opposite charge.  Attraction has also been
seen analytically
\cite{kjell_ake,oosawa,medina,patey,KM1,hnc1,kjell2,kjel4,lozada,clay,kjell_phys,kjell_pash,feller,great,marcel_membrane,att,stevens,tangzx,raymanning,rouzina,bj}
through various methods, including integral-equation
approaches,\cite{kjell_ake,medina,patey,KM1,hnc1,kjell2,kjel4,lozada,clay,kjell_phys,kjell_pash,feller,great,marcel_membrane,att}
density-functional theories,\cite{stevens,tangzx} and lattice
models.\cite{raymanning,rouzina} Finally, a variety of experiments
\cite{kjell_pash,bloom_cond,Bloom,isemore,ise3,ise,rau1,rau2,sedlak1,sedlak2,tang,khan,slade}
has indicated attractions as well.  For example, X-ray scattering
studies on colloidal particles made of poly(styrene sulfonate)
\cite{isemore,ise3} or silica \cite{ise} have shown phase separation
behavior, and when coupled with externally applied osmotic stress, has
shown DNA aggregation by multivalent ions \cite{rau1,rau2} and
ligands.\cite{rau1} Effective diffusion constants derived from light
scattering measurements on salt-free systems of poly(styrene
sulfonate) of varying molecular weights \cite{sedlak1} or of varying
concentrations \cite{sedlak2} have suggested macroion bundling.  Light
scattering combined with electron microscopy has shown the reversible
aggregation behavior of various filamentous biopolymers in the
presence of multivalent metal ions or peptide homopolymers.\cite{tang}
Deuterium-NMR and $^{23}$Na-NMR have been used to derive phase
diagrams of lamellar liquid crystalline systems.  \cite{khan} Finally,
in calcium clays the attractive forces between the mica surfaces have
been measured using the surface force apparatus,\cite{kjell_pash} and
the effects of osmolyte concentrations on the clays have been studied
by X-ray scattering.\cite{slade} Thus, there is substantial evidence
of effective attractions between charged objects of the same sign.


Theory has shown that ion-ion correlations (i.e., correlations in the
fluctuations of the ion positions) are the source of the attractions
in the numerically and analytically studied model systems.  Mean-field
descriptions, such as the Debye-H\"{u}ckel and Poission-Boltzmann
equations,\cite{gouychapman} which are often used to calculate forces
between macroions,\cite{brenner,brenner2,carnie,shulepov} always fail
to yield this attraction because they do not account for interparticle
correlations.\cite{kirk} On the other hand, integral equation
approaches naturally possess the framework for including such
interactions.  Fluctuation theories also can yield an attraction.
Oosawa\cite{oosawa} has shown (to dipole order) that the coupling of
longitudinal fluctuations in the counterionic atmospheres along two
parallel, like-charged, rod-like macroions produces an electrostatic
attraction.  Whereas Oosawa has predicted this attraction to increase
with temperature, we recently found by the use of Brownian dynamics
simulations that it increases as the temperature decreases.\cite{us}


Since the effective interaction between like-charged surfaces or two
like-charged rods has been fairly well studied, we will address here
the effective interaction between rod-like macroions and a
like-charged surface.  The geometry is motivated by experimental
systems involving highly negatively charged DNA molecules and
negatively charged membranes.  In fact, experimental methods have been
developed (for reviews, see Ref.\cite{hansma_review,lyub_rev}) to
adsorb DNA onto negatively charged substrates.  Among these methods is
the adhesion of DNA to mica surfaces by the use of multivalent cations
\cite{carlos,hansma3,hansma1,hender,vesenka,hansma2,guts,delain,zen,zen2}
or cationic detergents.\cite{shaper} The treatment of mica surfaces
with a solution of magnesium acetate, calcium acetate, or magnesium
chloride has been seen to stabilize adsorbed DNA and subsequently give
reproducible imaging by atomic force microscopy
\cite{carlos,hansma3,hansma1,hender,vesenka,hansma2,guts} or scanning
tunneling microscopy.\cite{delain,zen,zen2} These treatments are
believed to displace potassium ions in the mica with divalent cations
so as to provide stronger binding sites.  Similarly, divalent cations
have been shown to be more successful than monovalent cations in
binding DNA to quartz sand in DNA-degradation studies.\cite{roman}
Studies on gene-delivery systems have reported on the adsorption of
DNA to anionic liposomes with \cite{hagstrom} and without
\cite{cywang} the use of positively charged binding proteins, rather
than simple cations.  Finally, two-dimensional, fingerprint-like films
of DNA strands have been found to adhere to cationic bilayers
deposited onto untreated mica
substrates.\cite{yang_physB2,yang_physB1} Thus, experiments reveal
that adsorption of rod-like macroions onto like-charged substrates can
be mediated by oppositely charged ions, proteins, or membranes.


The model system we will investigate, as suggested the experiments
above, is an array of rigid, negatively charged rods parallel to a
negatively charged surface.  The question on which we focus is, can a
charged surface attract a like-charged rod?  Although the DNA in the
above experiments is admittedly flexible and the surfaces are not
necessarily strictly planar, we investigate this simpler system in
order to gain insight into the more complicated systems.  For
simplicity, we omit salt from our study in order to examine the
fundamental nature of the interaction between the like-charged
macroions.  Our method is to fix the array of rods at certain
distances from the surface, compute the time-averaged force on the
rods perpendicular to the surface using Brownian dynamics simulations,
and investigate the dependence of this force on a systematic variation
of model parameters, such as array height, macroion charge densities,
rod spacing, counterion valence, and rod size.  We also examine the
roles of electrostatic images and the discretization of the surface
charge into mobile monovalent anions.  In the next section we present
the pairwise potential energy functions used in the simulations.
These potentials model only the classical electrostatic interactions
and neither distinguish between counterion species of the same valence
nor consider solvent structure.  We then summarize the Brownian
dynamics algorithm by which the representations of the counterions are
moved by deterministic and thermal forces and follow with the results
and a discussion.

\section{Theory}
\subsection{The Model}

The system we consider is an isotropic, continuous dielectric medium
containing negatively charged rods and their counterions above a
supported bilayer of anionic lipids.  Our simulational unit cell model
(Fig.~1) is a rectangular open half-space $(z>0)$ of dielectric
constant $\epsilon_1$ ($\approx 80$ for an aqueous medium) with
dimensions $L_x$ and $L_y$ that contains a single line charge of
uniform charge density $\lambda$ located a distance $d$ from a fixed,
flat surface of a specified average charge density $\sigma$.  This
surface density can be either a uniform density distribution or a
collection of discrete, mobile monovalent anions, representing, e.g.,
a fluid-phase mixture of charged and neutral lipids.  The unit cell is
replicated the $x$ and $y$ directions to produce an infinite array of
infinitely long, parallel line charges with a spacing $L_x$ and with a
repeat distance $L_y$ in the $y$ direction.  There is no periodicity
in the $z$ direction.  This replicated dielectric medium is bounded
below ($z \leq 0$) by an isotropic, continuous half-space of
dielectric constant $\epsilon_2$ ($\approx 3$) representing the
dielectric layer formed by the hydrocarbon tails of the lipids.  The
dielectric discontinuity at $z=0$ produces electrostatic images in the
substrate in order to satisfy the boundary conditions on the
dielectric displacement field at the interface.\cite{jackson} A point
particle of charge $q_i$ placed in the medium with $\epsilon_{1}$ at a
distance $z_i$ away from the planar interface will produce an
electrostatic image charge of charge $q^\prime_i = q_{i} (\epsilon_{1}
- \epsilon_{2})/(\epsilon_{1} + \epsilon_{2})$ at the symmetric
position $-z_i$.  An electrostatic image rod is produced similarly.
As a result, a uniform surface charge density (or the charge of
individual mobile monovalent surface anions) increases by a factor of
$2\epsilon_1/(\epsilon_1 + \epsilon_2)$.

The interaction potentials consist of pairwise, long-ranged
electrostatic (Coulombic) forces and short-ranged, non-electrostatic
repulsions.  We make no approximations regarding the Coulomb
interations.  For a uniformly charged surface, the ion-surface
electrostatic interaction $V_{\rm i-s}(z)$ is a function of the
distance $z$ the ion with charge $q_i$ is from the surface:
\FL\begin{eqnarray}
\quad V_{\rm i-s}(z) = -{q \sigma \over 2 \epsilon_1\epsilon_0} z,
\end{eqnarray}
where $\epsilon_0$ is the permittivity of vacuum.
The rod-surface electrostatic contribution $V_{\rm r-s}(z)$ 
per unit cell for a uniform surface charge distribution is similarly
\FL\begin{eqnarray}
\quad V_{\rm r-s}(z) = -{\lambda L_y \sigma \over 2 \epsilon_1\epsilon_0} z.
\end{eqnarray}
The ion-ion electrostatic potential energy $V_{\rm i-i}(\Delta{\bf
r})$ for an ion with charge $q_1$ at the point $(x + \Delta x, y +
\Delta y, z + \Delta z)$ and an ion with charge $q_2$ and its replicas 
located at $(x+mL_x,y+nL_y,z)$, where $m,n$ are integers, resulting 
from the two-dimensional replication of the unit cell in the $x$ and 
$y$ directions, is given {\it exactly} by an absolutely convergent, 
double summation.  The ion-ion potential energy 
$V_{\rm i-i}(\Delta{\bf r})$ is\cite{ngj_1oR} 
\FL
\begin{eqnarray}
\quad V_{\rm i-i}(\Delta{\bf r}) = {q_1 q_2\over 
4\pi\epsilon_1\epsilon_0}\Biggl\{ {4 \over L_x }\sum^\infty_{n=1} 
\cos\left(2\pi{\Delta x\over L_x}n\right) \sum^\infty_{k=-\infty} \Biggr. \nonumber 
\end{eqnarray}
\vspace{-1.5\baselineskip}
\FL\begin{eqnarray}
\qquad K_0\left\{ 
2n\pi n \left[ 
\left({L_y\over L_x}\right)^2 \left({\Delta y\over L_y}+k\right)^2 +
\left({\Delta z\over L_x}\right)^2
\right]^{1/2} 
\right\} \nonumber
\end{eqnarray}
\vspace{-1.5\baselineskip}
\FL\begin{eqnarray}
\qquad  -{1\over L_x}\ln\left[\cosh\left(2\pi{\Delta z\over L_y}\right) -
\cos\left(2\pi{\Delta y\over L_y}\right)   \right] \nonumber 
\end{eqnarray}
\vspace{-1.5\baselineskip}
\FL\begin{eqnarray}
\label{eqn_vii}
\qquad \Biggl. -{\ln 2 \over L_x} \Biggr\}.
\end{eqnarray}
When the surface consists of discrete ions, the potentials 
$V_{\rm i-s}(z)$ and $V_{\rm r-s}(z)$ are identically zero, and 
all electrostatic interactions with the surface are taken according
to Eq.~(\ref{eqn_vii}).
The self-energy $V_{\rm i-i}^{(0)}$ that arises from an ion
interacting
with its own replicas is found by evaluating the expression
\FL
\begin{eqnarray}
\quad V_{\rm i-i}^{(0)} = {1\over 2}\lim_{r\rightarrow 0} 
   \left( V_{\rm i-i}(\Delta{\bf r}) - {q_i^2\over 4\pi\epsilon_1\epsilon_0
r}\right)
\end{eqnarray}
and is given in Ref.\cite{ngj_1oR} The ion-rod potential energy
$V_{\rm i-r}(\Delta{\bf r})$ is the logarithmic interaction between a
point particle and a one-dimensional array of line charges.  The
analytic form of $V_{\rm i-r}(\Delta{\bf r})$ is derived from the
potential energy of an ion interacting logarithmically with a
two-dimensional array of line charges arranged on a rectangular
lattice (see Eq.~(14) in Ref.\cite{ngj_logR}) by eliminating one of
the dimensions.  The result is given by
\FL\begin{eqnarray}
\quad V_{\rm i-r}(\Delta{\bf r}) = -{q\lambda\over 4\pi\epsilon_1\epsilon_0} 
   \ln \left\{ 2 \left[ \cosh\left(2\pi{\Delta z\over L_x}\right) \right. \right. \nonumber 
\end{eqnarray}
\vspace{-1.5\baselineskip}
\FL\begin{eqnarray}
\label{logint}
\label{eqn_vir}
\qquad\qquad\qquad \left. \left. - \cos\left(2\pi{\Delta x\over L_x}\right)
\right] \right\}.
\end{eqnarray}
The corresponding ion-rod self-energy simplifies similarly 
from Ref.\cite{ngj_logR} to
\FL\begin{eqnarray}
\quad V_{\rm i-r}^{(0)} = -{q\lambda\over 4\pi\epsilon_1\epsilon_0}\ln{2\pi\over L_x}.
\end{eqnarray}
The interaction between the real ions and electrostatic image ions as
well as between the real ions (rods) and electrostatic image rods
(ions) is obtained directly from the periodic potentials in
Eqs.~(\ref{eqn_vii}--\ref{eqn_vir}), respectively, up to a factor of
${2\epsilon_1/(\epsilon_1+\epsilon_2)} \approx 1.93$).  Thus,
interactions involving electrostatic images are treated {\it without}
approximation.  Finally, the ion-rod and ion-surface short-ranged,
non-electrostatic repulsion energies were modeled as $A_{\rm
  i-r}/r^{11}$ and $A_{\rm i-s}/r^{10}$, respectively, to prevent the
ions from collapsing electrostatically.  The combination of the
Coulomb interaction and short-ranged repulsion yields an optimal
ion-rod distance, i.e., a rod radius $r_0$, which was considered among
the model parameters.  Values for the coefficients $A_{\rm i-r}$ and
$A_{\rm i-s}$ and the resulting optimal ion-rod and ion-surface
distances are given below.  The ion-rod short-ranged interaction is
taken according to the minimum image convention.

The values of the parameters in our reference system are taken to be
reasonably representative of those found in the literature.  Under
physiological conditions the DNA found in most organisms
is\cite{stryer} in the so-called B-form, a double-helical structure of
approximate radius of 12\,{\AA}, bearing two negatively charged
phosphate groups per base pair (i.e., every 3.4\,{\AA} on average).
Taken as a whole, calcium clays are reported\cite{clay} to have areas
per unit charge ranging from 51--135\,\AA$^2$, whereas negatively
charged lipids known\cite{lipids} to form (mono/bilayers) have a
median surface area per molecule of about 51\,{\AA}$^2$.  On the basis
of this latter value alone, we choose our reference system to consist
of rods with an equivalent linear charge density $\lambda_{\rm ref} =
-e/1.7$\,{\AA}, lateral spacing $L_x = 30$\,{\AA}, and a surface with
an average charge density $\sigma_{\rm ref} = -e/51$\,{\AA}$^2$.  As
we were unable to find data on the spacing of DNA in fully
two-dimensional films adsorbed directly onto negatively charged
substrate, we do not know whether a lateral spacing of 30\,{\AA} is
reasonable.  However, DNA molecules adsorbed onto cationic-lipid
bilayers supported by mica have\cite{yang_physB1} an interhelical
spacing that depends nonmonotonically on the concentration of divalent
magnesium cations in solution and lies in the range 35--45\,{\AA}.
Furthermore, DNA molecules in sodium chloride solution are
known\cite{livolant,podgornik} to pack into hexagonal arrays with
lattice spacings ranging between 25--40\,{\AA}.  Dielectric constants
in our system are taken to be $\epsilon_1 = 80$ and $\epsilon_2 = 3$
or 80.  In the former case, when electrostatic images are included, an
ion or rod always repels its own electrostatic image.  All simulations
were carried out at 300~K.

Values of $A_{\rm i-r}$ of $7.5\times 10^7$, $3.27\times 10^4$, and
$1.8\times 10^1$ kcal \AA$^{11}$/mol yielded rod radii $r_0$ of 5.6,
2.8, and 1.4 \AA, respectively, in the reference macroion system.
Unless otherwise specified, the second $A_{\rm i-r}$ value was chosen
as the reference rod value regardless of the actual macroion charge
densities.  For all simulations $A_{\rm i-s}$ was fixed at $6.63\times
10^2$ kcal \AA$^{10}$/mol, yielding an optimal ion-surface distance of
2.4\,{\AA} for the reference system.  The reference value of rod
radius $r_0 = 2.8\,{\rm\AA}$ is not equal to the real size of DNA. The
reason for this is that the rod size and the correct ion-rod
interaction energy cannot be imposed simultaneously within our model.
We believe it would be more correct to account for the electrostatic
``binding'' energy between a counterion and a DNA ion (on the
order\cite{nacl} of a few $k_{\rm B}T$) rather than the actual DNA
size.

\subsection{Simulation Method}

The motion of a collection of particles undergoing Brownian dynamics
corresponds to the long-time behavior of the Langevin equation in the
overdamped limit.\cite{allen}  The Langevin equation expresses a
balance of deterministic and thermal, random forces on each particle
at any given moment in time $t$.  For each particle $i$ with mass
$m_{i}$, position ${\bf r}_i(t)$, velocity ${\bf v}_i(t)$, and
isotropic friction coefficient $\xi_{i}$, the Langevin equation reads
\FL
\begin{eqnarray}
\label{sim1}
\quad m_i {d{\bf v}_i(t)\over dt} = -\xi_i m_{i}{\bf v}_i(t) + {\bf 
f}_i(\{{\bf r}(t)\}) + 
{\bf f}_i^*(t)
\end{eqnarray}
where ${\bf f}_i(\{{\bf r}(t)\})$ is the total deterministic force
acting on particle $i$ due to the positions $\{{\bf r}(t)\}$ of all
the particles, and ${\bf f}_i^*(t)$ is the random force acting on particle
$i$.  This random force is characterized by the 
fluctuation-dissipation theorem as
\FL\begin{eqnarray}
\label{sim2}
\quad\langle {\bf f}_i^*(0)\cdot {\bf f}_j^*(t) \rangle = 
6 m_i k_{\rm B} T \xi_i\/ \delta(t)\delta_{ij}
\end{eqnarray}
for particles $i$ and $j$, with the time average of the random force 
over any particle vanishing:
\FL\begin{eqnarray}
\quad\langle {\bf f}_i^*(t)\rangle = 0.
\end{eqnarray}
At long times the friction coefficent $\xi_i$ is related to the
diffusion constant $D$ as $D_i = k_{\rm B} T/m_{i}\xi_i$.  The
inertial term on the left-hand side of Eq.~(\ref{sim1}) can be set to zero at
long times.  One discretization of the resulting equation
yields an iterative scheme for the time evolution of the system of
particles:\cite{yeh,ermak1}
\FL\begin{eqnarray}
\label{sim3}
\quad{\bf r}_i (t + \Delta t) = {\bf r}_i (t) + {\Delta 
t\over m_i\xi_i}{\bf f}_i(\{{\bf r}(t)\}) + {\bf r}^*_i(\Delta t)
\end{eqnarray}
where ${\bf r}^*_i(\Delta t)$ is a random Gaussian displacement with
a variance of $2 k_{\rm B} T \Delta t /m_i \xi_i$ in each dimension.
Given the total potential energy surface for a particle $i$,
$V_{i,{\rm tot}}(\{{\bf r}(t)\})$, the deterministic force ${\bf
f}_i(\{{\bf r}(t)\})$ is given by ${\bf f}_i(\{{\bf r}(t)\}) =
-\nabla V_{i,{\rm tot}}(\{{\bf r}(t) \})$.

\section{Results and Discussion}

Each simulation required the specification of a set of system
parameters $\{\lambda, \sigma, r_0, L_x, L_y\}$, a fixed position of
the rods $d$, whether the surface charge density consisted of
discrete, mobile monovalent anions, and whether electrostatic images
were to be included.  The value of $L_y$ depended on the particular
set of system parameters and fell in the range 17--272\,{\AA}, thereby
giving approximately 80 or more ions in the basic unit cell to help
ensure that particle configurations were well sampled.  The divalent
or monovalent counterions were introduced as a random configuration
until the system became overall charge neutral; no additional ions
(i.e., co-ions or salt) were added.  As the rods were fixed in space,
the simulation involved only the motion of the ions due to the
explicit pairwise interactions among the ions, the rods, and the
single charged surface, as well as the random thermal forces.  For the
ion motion, the normalized time step was taken to be $\Delta t \leq
0.01$ for $m_i \xi_i \equiv 1$.  During the simulations the vertical
and lateral time-averaged forces per unit length on the rod and the
vertical time-averaged force per unit $L_y$ length on the surface were
monitored.  After the initial 5--10\,$\times 10^4$ steps were
discarded, these averages were accumulated over an additional
2--200\,$\times 10^6$ time steps, depending on the number of simulated
particles, system size, macroion charge densities, and counterion
valences.  Equilibrium was attained when the average force on the rod
$\langle F_{z,{\rm rod}} \rangle$ and surface 
$\langle F_{z,{\rm surf}} \rangle$ remained 
constant and the average lateral force on
the rod $\langle F_{z,{\rm rod}} \rangle \approx 0$.  We checked for
finite-size effects in $L_y$ and found that they do not affect our
results significantly.

Numerous simulations with systematic variations of the system
parameters were carried out to investigate the general behavior of the
system.  Figure~2 shows the time-averaged, rod-surface force in the
$z$-direction per unit length rod as a function of rod-surface spacing
for several rod and uniform surface charge densities for monovalent
and divalent counterions, with and without electrostatic images.  For
the case of monovalent counterions, the top set of two curves shows
that the effective rod-surface interaction is always repulsive,
whereas the lower set of curves indicates that there are rod-surface
distances $d$ for which the interaction is attractive.  As can be
seen, the interaction becomes more attractive as the charge densities
on the rod and surface increase from one to four times the reference
charge densities.  For the case of divalent counterions, the
rod-surface interaction is attractive for half, one times, and double
the reference charge densities.  The maximal (attractive) force,
indicated by the depth of the force-distance profile, increases as the
charge densities increase.  Divalent counterions are obviously more
effective at inducing an attraction than are monovalent counterions.

The effect of changing the interrod spacing $L_x$ on a
divalent-counterion system at the reference rod and uniform surface
charge densities was examined.  For a system containing no
electrostatic images, the force-distance profiles for $L_x = 15, 30,
60$\,{\AA} were indistinguishable, and the profiles for systems with
electrostatic images were also indistinguishable.  Halving (doubling)
the rod radius $r_0$ also resulted in indistinguishable profiles whose
maximal attractive forces, i.e., minima, were deeper (shallower) and
located at smaller (larger) $d$ compared to the reference case.  Thus,
these combinations of $L_x$ and $r_0$ place these systems essentially
in the dilute-rod limit.  This dilute-rod regime is a well-defined
limit as the effective rod-surface interaction is independent of the
initial distribution of counterions.  In further decreasing the
lateral rod spacing $L_x$ (and/or increasing the rod radius) one would
eventually encounter a less permeable array of rods.  The resulting
effective rod-surface force is then expected to depend longer on time
and may lead to a significantly different effective rod-surface force
as compared to the dilute-rod regime, due to the initial partitioning
of counterions between the region exterior to the rods ($z>d$) and the
region between rods and surface ($0 < z < d$, in Fig.~1).

Figure~2 can also be interpreted in terms of the equilibrium
rod-surface separation distance, $d^*$, the point at which
force-distance profiles cross zero.  In general, for either counterion
species, as the macroion charge densities increase, the equilibrium
rod-surface distance decreases.  This behavior is most striking for
monovalent counterions as increasing the macroion charge densities
causes $d^*$ to move from infinity to a finite distance.  Finally, the
effect of electrostatic images for both monovalent- and
divalent-counterion systems is quite small compared to that which can
be achieved through changing only the macroion charge densities.

Changing the macroion charge densities individually leads to a type of
phase diagram for the rod-surface force.  Figure~3 shows the regions
of rod-surface attraction and repulsion for systems of monovalent
counterions with uniform charge densities and no electrostatic images.
In general, the larger macroion charge densities (on the rod or the
surface) lead to an effective attraction, and the lower charge
densities lead to repulsion.  The parameter set $\lambda/\lambda_{\rm
  ref} = \sigma/\sigma _{\rm ref} = 2$ led to only marginal
attraction, and this ambiguity is indicated in the figure.  The dashed
line, as suggested by the data, indicates only a qualitative phase
boundary between the attractive and repulsive regions, as the actual
shape of the boundary is unknown and would require a more detailed
exploration of the macroion density phase space.  Doubling the rod
radius led to repulsion in each of the nine cases shown in Fig.~3.
For a given fixed rod-surface separation distance, an increase in the
rod radius tends to obstruct the counterions from entering the region
in between the rod and surface, thereby hindering the ions from
contributing relevant attractive interactions.  Thus, the presence of
an attraction bewteen the rods and surface is conditional on not only
the macroion charges but also the rod radius.

We examined the effect of macroion charge densities in systems of
divalent counterions and found, as in Fig.~1, that an attraction
between the rods and surface is evident in systems with macroion
uniform charge densities down to $\lambda = 0.5\lambda_{\rm ref}$ and
$\sigma = 0.5\sigma_{\rm ref}$ with the reference rod size.  In these
cases, the equilibrium rod-surface distance lies within a quite small
range (5--6\,{\AA}), with the smaller distances corresponding to the
higher charge densities.  We have not checked whether an attraction
exists at even lower charge densities on either macroion.

Finally, the roles of a discrete, mobile surface charge distribution
and electrostatic images in systems of monovalent and divalent
counterions were examined.  Figure~4 shows a comparison of systems
with the reference densities $\lambda = \lambda_{\rm ref}$ and $\sigma
= \sigma_{\rm ref}$ for the cases of 1) a uniformly charged surface
with no electrostatic images, 2) a uniformly charged surface with
electrostatic images, 3) a surface composed of discrete, mobile
monovalent anions with no electrostatic images, and 4) same as in 3)
but with electrostatic images.  The force-distance profiles for the
monovalent-counterion systems all show a repulsive rod-surface force
and are essentially indistinguishable.  Although the divalent
force-distance profiles give nearly the same equilibrium rod-surface
distances and appear similar in shape, the depths of the profiles vary
significantly.  The maximal attractive force on the rods is greater
for systems with surfaces composed of discrete, mobile charges than
for systems with a uniformly charged surface.  It is reasonable to
attribute this result of enhanced attractions to the increased
coupling between the mobile surface ions and the rod and other ions.
A surface composed of a mobile charge distribution provides the system
with more configurational degrees of freedom, which is, in general,
expected to lead to an overall lowering of the free energy of the
system.  Interestingly, the four force profiles for the
divalent-counterion systems all cross near 7\,{\AA}.  We do not know
whether this common point is a general phenomenon or an artifact due
to our particular system geometry and parameters.

\section{Conclusions}

In summary, we have presented a systematic study of the effective
interactions between a single charged surface and a parallel array of
same-sign charged rods in the presence of their counterions with no
co-ions (salt).  We indeed find an effective attraction between the
rods and surface that is mediated by either divalent or monovalent
counterions at room temperature in the dilute-rod limit.  This
attraction results from counterion correlations that overcome the
inherent repulsion between the two macroions.  Divalent counterions
readily mediate this attraction at macroion charge densities that are
comparable to those found in highly charged biopolymers and surfaces.
Clearly, counterions of higher valence are more effective at inducing
attraction between charged macroions of the same-sign charge density,
and this trend has been observed experimentally \cite{Bloom,tang} and
numerically.\cite{guldnils,us} Although the very high macroion charge
densities needed by systems of only monovalent counterions to display
an effective attraction are seemingly unphysical, there indeed appears
to be polymers with rather high equivalent linear charge densities.
One group\cite{szoka} has pointed out that the anionic polymers
dextran sulfate and heparin have a 2.0-fold greater charge density
than DNA.  The simulations have further revealed that the higher the
macroion charge densities, the smaller the rod-surface equilibrium
distance.

We have demonstrated for this rod-surface system that the volume
excluded by the rod to the counterions is an important factor in
determining macroion attraction or repulsion, particularly for systems
of monovalent counterions.  An increase in rod radius tends to lower
the attraction between the macroions because the average number of
counterions in between the macroions is reduced.  These results are
consistent with our recent work \cite{us} on systems of two identical
rods in the absence of salt.  In the present paper we also have
discussed that the decrease in the maximal attraction due to an
increased rod size can be partially offset by an increase in the
rod-surface distance, while leading to a new decreased maximal
attractive force.

The accounting for electrostatic images generally introduces quite
small, repulsive additions to the rod-surface interaction.  Whereas we
have limited the counterions from coming within about 2.4\,{\AA} from
the surface, further investigation revealed that the inclusion of
electrostatic images while decreasing the optimal ion-surface distance
to about 0.5\,{\AA} can increase the rod-surface equilibrium distance
by 1\,{\AA} compared to the case without electrostatic images.  Thus,
the inclusion of electrostatic images in these systems seems to
introduce only rather short-ranged, repulsive effects.  This raises a
question for real aqueous systems and their modeling: To what extent do
the hydration layers around macroions conceal electrostatic image
effects?

There are many variations of this system that one could explore.  For
example, in the present system the effective rod-surface force could
be studied as a function of the ratio of monovalent and divalent
counterions.  As the strengths of the electrostatic interactions
involving monovalent counterions would be less than those involving
divalent counterions, the ``dilution'' of a divalent-counterion system
by the addition of monovalent counterions would be expected to
decrease the attractive force between the rod and surface.  Another
variation of the present system geometry that has immediate relevance
to experiments is a system consisting of {\it oppositely} charged rods
and surface.  In particular, Yang and Fang \cite{yang_physB2} imaged
DNA strands adsorbed onto a supported cationic lipid bilayer using
atomic force microscopy and found that the DNA spacing in the
fingerprint-like images increased monotonically from about 40 to
60\,{\AA} as the sodium chloride concentration increased from about
20\,mM to 1\,M.  This latter subject will be addressed in a future
article.

\section*{Acknowledgments}

We thank Robijn Bruinsma, William Gelbart, and Philip
Pincus for stimulating discussions.  This work was performed under the
auspices of the United States Department of Energy and was supported
in part by NSF grant DMR--9708646.


\begin{figure}[h]\label{fig1}
\begin{center}
\epsfig{file=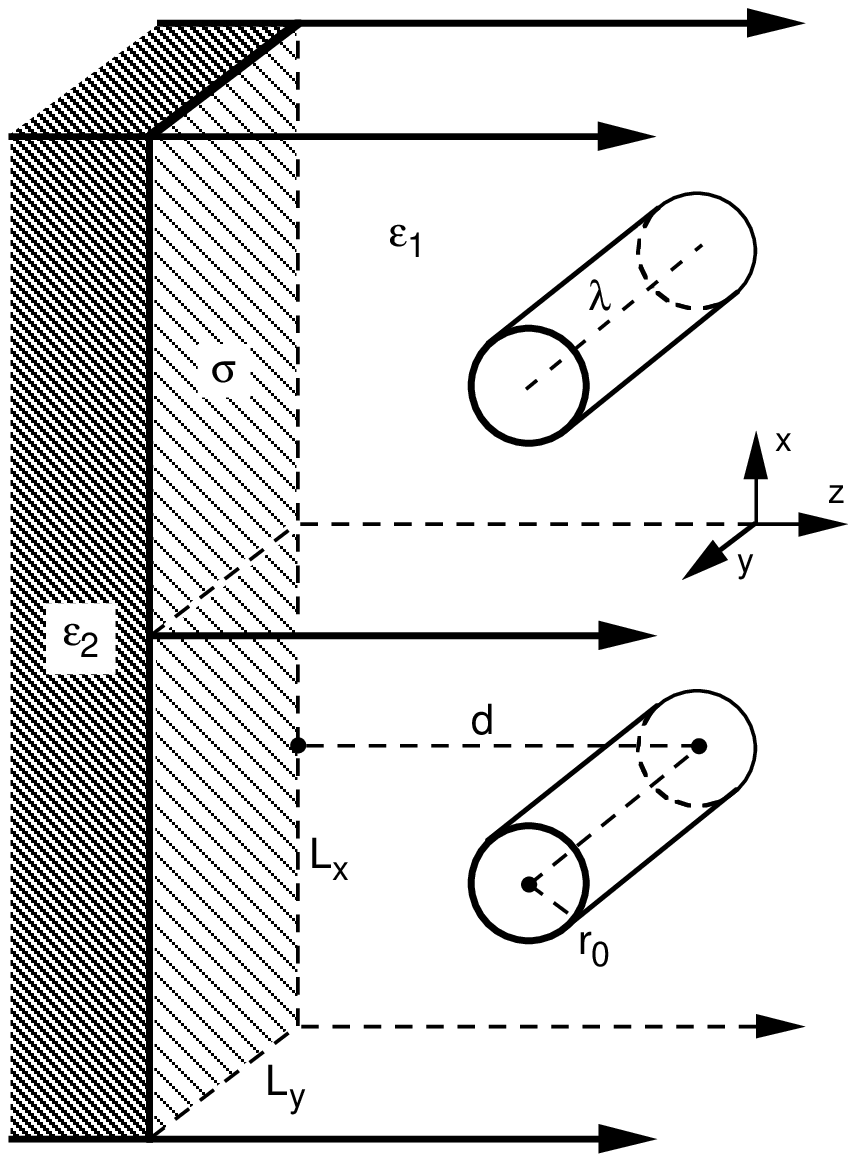}
\end{center}
\caption{
  The simulation system as described in the text.  Periodic boundary
  conditions applied to a unit cell of lateral dimensions $L_x$ and
  $L_y$ produce an infinite array of infinitely long line charges with
  a spacing of $L_x$.  Each line charge with uniform charge density
  $\lambda$ and radius $r_0$ is located in an isotropic, continuous
  medium of dielectric constant $\epsilon_1$ and positioned a distance
  $d$ from a charged surface at $z=0$ with average density $\sigma$.
  The replicated $\epsilon_1$ medium is bounded by a second isotropic,
  continuous medium of dielectric constant $\epsilon_2$.
  }
\end{figure}

\begin{figure}[h]\label{fig2}
\begin{center}
\epsfig{file=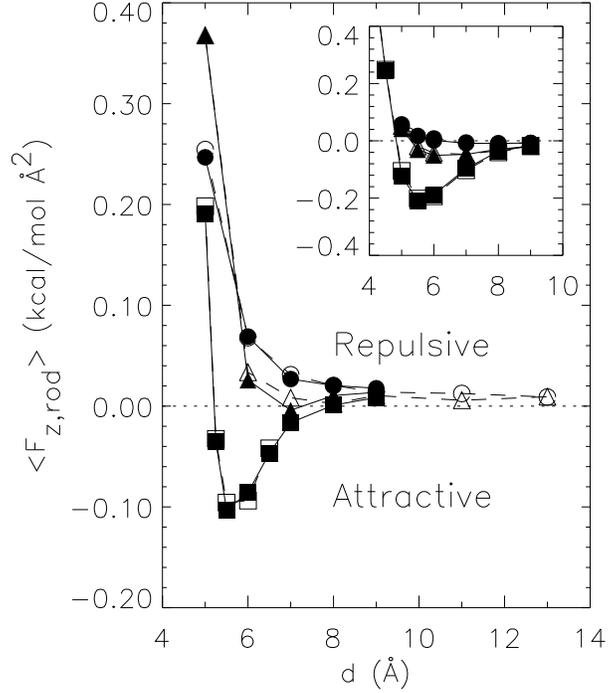}
\end{center}
\caption{
  Force-distance profiles for the effective, time\-averaged force on
  the rod (per unit length) perpendicular to the surface for a given
  fixed rod-surface distance $d$.  Shown are a series of curves for a
  system of monovalent counterions with (dashed lines, open symbols)
  and without (solid lines, filled symbols) electrostatic images.  The
  symbols correspond to the following uniform charge densities:
  $(\bullet,\circ)~ \lambda = \lambda_{\rm ref}, \sigma = \sigma _{\rm
    ref}$; $(\blacktriangle,\vartriangle)~\lambda = 2\lambda_{\rm
    ref}, \sigma = 2\sigma _{\rm ref}$; $(\blacksquare,\square)~
  \lambda = 4\lambda_{\rm ref}, \sigma = 4\sigma _{\rm ref}$.  The
  inset shows the analogous series for a system of divalent
  counterions with the following charge densities: $(\bullet,\circ)~
  \lambda = 0.5\lambda_{\rm ref}, \sigma = 0.5\sigma _{\rm ref}$;
  $(\blacktriangle,\vartriangle)~\lambda = \lambda_{\rm ref}, \sigma =
  \sigma _{\rm ref}$; $(\blacksquare,\square)~ \lambda = 2\lambda_{\rm
    ref}, \sigma = 2\sigma _{\rm ref}$.  The sizes of the error bars
  do not exceed those of the figure symbols and are not shown.  In all
  cases, $L_x = 30$\,{\AA}, $r_0 \approx 2.8$\,{\AA}, $\lambda_{\rm
    ref} = -e/1.7$\,{\AA}, and $\sigma_{\rm ref} = -e/51$\,{\AA}$^2$.
  }
\end{figure}

\begin{figure}\label{fig3}
\begin{center}
\epsfig{file=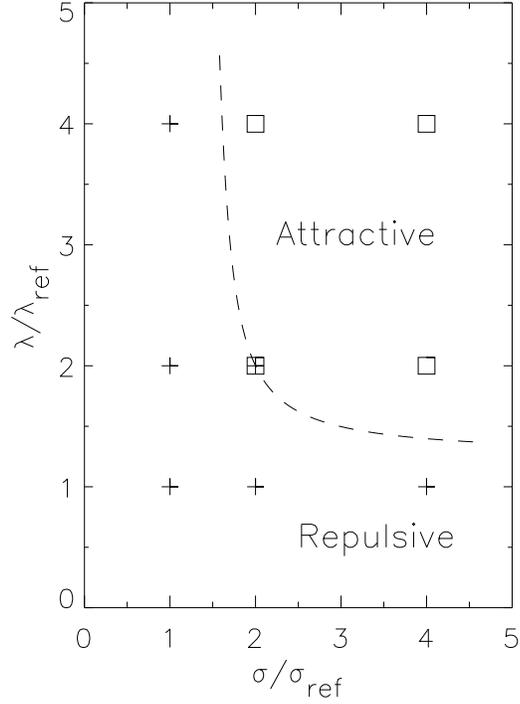}
\end{center}
\caption{
  Phase diagram for a system of monovalent counterions, indicating
  regions in which the rod-surface interaction is attractive $(\Box)$
  and repulsive $(+)$ for several values of uniform rod linear charge
  densities $\lambda$ and uniform surface charge densities $\sigma$,
  relative to the reference system with $\lambda_{\rm ref}$ and
  $\sigma _{\rm ref}$. The point at $\lambda/\lambda_{\rm ref} =
  \sigma/\sigma _{\rm ref} = 2$ corresponds to a system that shows
  only marginal attraction (see Fig.~2) and is marked to indicate that
  it is a boundary point.  The dashed line indicates qualitatively the
  boundary between the regions. In all cases, $L_x = 30$\,{\AA}, $r_0
  \approx 2.8$\,{\AA}, $\lambda_{\rm ref} = -e/1.7$\,{\AA} and
  $\sigma_{\rm ref} = -e/51$\,{\AA}$^2$.  }
\end{figure}

\begin{figure}\label{fig4}
\begin{center}
\epsfig{file=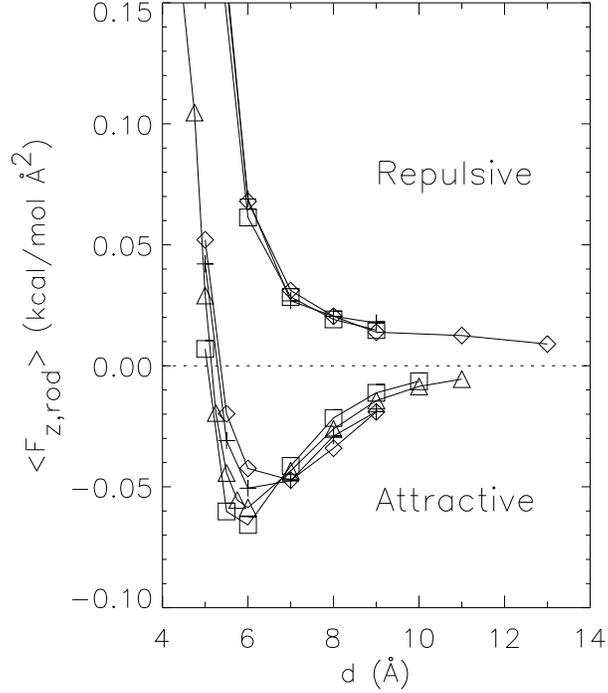}
\end{center}
\caption{
  Comparison of systems with or without electrostatic images, having
  surfaces with either uniform charge densities or discrete mobile
  monovalent anionic charge distributions for $\lambda = \lambda_{\rm
    ref}$ and $\sigma = \sigma _{\rm ref}$.  The upper curves
  correspond to systems of monovalent counterions; the lower curves,
  divalent.  The following key applies to both sets of curves: $(+)$
  uniform with no electrostatic images, $(\Diamond)$ uniform with
  electrostatic images, $(\triangle)$ discrete mobile surface with no
  electrostatic images, $(\Box)$ discrete mobile surface with
  electrostatic images.  The sizes of the error bars do not exceed
  those of the figure symbols and are not shown.  In all cases, $L_x =
  30$\,{\AA}, $r_0 \approx 2.8$\,{\AA}, $\lambda_{\rm ref} =
  -e/1.7$\,{\AA} and $\sigma_{\rm ref} = -e/51$\,{\AA}$^2$.
  }
\end{figure}

\end{document}